\documentclass[structabstract]{aa}  
\usepackage{epsfig}
\usepackage{natbib}
\usepackage{txfonts}
\begin{document}
\newcommand{\bea}{\begin{eqnarray}}    
\newcommand{\eea}{\end{eqnarray}}      
\newcommand{\be}{\begin{equation}}
\newcommand{\ee}{\end{equation}}
\newcommand{\bef}{\begin{figue}}
\newcommand{\eef}{\end{figure}}
\newcommand{\etal}{et al.}
\newcommand{\kms}{\,{\rm km}\;{\rm s}^{-1}}
\newcommand{\hubunits}{\,\kms\;{\rm Mpc}^{-1}}
\newcommand{\hmpc}{\,h^{-1}\;{\rm Mpc}}
\newcommand{\hkpc}{\,h^{-1}\;{\rm kpc}}
\newcommand{\msun}{M_\odot}
\newcommand{\K}{\,{\rm K}}
\newcommand{\cm}{{\rm cm}}
\newcommand{\cd}{{\langle n(r) \rangle_p}}
\newcommand{\Mpc}{{\rm Mpc}}
\newcommand{\kpc}{{\rm kpc}}
\newcommand{\xir}{{\xi(r)}}
\newcommand{\xrp}{{\xi(r_p,\pi)}}
\newcommand{\xsirpi}{{\xi(r_p,\pi)}}
\newcommand{\wrp}{{w_p(r_p)}}
\newcommand{\gr}{{g-r}}
\newcommand{\Navg}{N_{\rm avg}}
\newcommand{\Mmin}{M_{\rm min}}
\newcommand{\fiso}{f_{\rm iso}}
\newcommand{\Mr}{M_r}
\newcommand{\rp}{r_p}
\newcommand{\zmax}{z_{\rm max}}
\newcommand{\zmin}{z_{\rm min}}

\def\eg{{e.g.}}
\def\ie{{i.e.}}
\def\spose#1{\hbox to 0pt{#1\hss}}
\def\ltapprox{\mathrel{\spose{\lower 3pt\hbox{$\mathchar"218$}}
\raise 2.0pt\hbox{$\mathchar"13C$}}}
\def\gtapprox{\mathrel{\spose{\lower 3pt\hbox{$\mathchar"218$}}
\raise 2.0pt\hbox{$\mathchar"13E$}}}
\def\inapprox{\mathrel{\spose{\lower 3pt\hbox{$\mathchar"218$}}
\raise 2.0pt\hbox{$\mathchar"232$}}}

\title{Absence of anti-correlations and of baryon acoustic
  oscillations in the galaxy correlation function from the Sloan
  Digital Sky Survey DR7}

\subtitle{}

\author{Francesco Sylos Labini \inst{1,2} \and Nikolay L. Vasilyev
  \inst{3} \and Yurij V. Baryshev \inst{3} \and Mart\'in
  L\'opez-Corredoira \inst{4}}

\titlerunning{Absence of anti-correlations and of BAO in the SDSS-DR7} 
\authorrunning{Sylos Labini et al.}
\institute{ 
Centro Studi e Ricerche Enrico Fermi, Via Panisperna 89 A, 
Compendio del Viminale, 00184 Rome, Italy
\and Istituto dei Sistemi Complessi CNR, 
Via dei Taurini 19, 00185 Rome, Italy.
\and 
Institute of Astronomy, St.Petersburg 
State University, Staryj Peterhoff, 198504,
St.Petersburg, Russia
\and
Instituto de Astrof\'isica de Canarias
C/.V\'ia L\'actea, s/n
ES-38200 La Laguna, Tenerife (Spain)}

\date{Received / Accepted}

\abstract 
{} 
{One of the most striking features predicted by standard models of
  galaxy formation is the presence of anti-correlations in the matter
  distribution on large enough scales ($r>r_c$). Simple arguments show
  that the location of the length scale $r_c$, marking the transition
  from positive to negative correlations, is the same for any class of
  objects as for the full matter distribution; i.e. it is invariant
  under biasing. This scale is predicted by models to be at about the
  same distance of the scale signaling the baryonic acoustic
  oscillation scale $r_{bao}$.  }
{ We test these predictions in the newest
  SDSS galaxy samples where it is possible to measure correlations
  on $\sim 100$ Mpc/h scales both in the main galaxy (MG) and
  in the luminous red galaxy (LRG) volume-limited samples. We
  determine, by using  three different estimators, the redshift-space
  galaxy two-point correlation function.}
{ We find that, in several MG samples, the correlation function
  remains positive on scales $>250$ Mpc/h, while it should be negative
  beyond $r_c\approx 120$ Mpc/h in the concordance LCDM.  In other
  samples, the correlation function becomes negative on scales $<50$
  Mpc/h.  To investigate the origin of these differences, we considered
  in detail the propagation of errors on the sample density into the
  estimation of the correlation function.  We conclude that these are
  important at large enough separations and that they are responsible
  for the observed differences between different estimators and for
  the measured sample-to-sample variations in the correlation
  function. We show that in the LRG sample the scale corresponding to
  $r_{bao}$ cannot be detected because fluctuations in the density
  fields are too large in amplitude.  Previous measurements in similar
  samples have underestimated volume-dependent systematic effects.
} 
{We conclude that, in the newest SDSS samples, the large-scale
  behavior of the galaxy correlation function is affected by intrinsic
  errors and volume-dependent systematic effects that  make the
  detection of correlations only an estimate of a lower limit of
  their amplitude, spatial extension, and statistical errors.  We point
  out that these results represent an important challenge to LCDM
  models as they largely differ from its predictions.
}

\keywords{Cosmology: observations; large-scale structure of Universe; } 
\maketitle


\section{Introduction}

Standard models of galaxy formation (i.e., cold, warm and hot dark
matter models) predict the two-point correlation function $\xi(r)$ of
matter density fluctuations in the early universe, and they can make a
simple prediction for that at the present time, in the regime of weak
density perturbations, where fluctuations have been only linearly
amplified by gravitational clustering in the expanding universe
\citep{pee80}.
 The difference in the various models lying in the values of the
 characteristic length scales and in the particular scale-behavior of
 $\xi(r)$. In general, this is characterized by three length scales
 and three different regimes,
(i) on scales smaller than $r_0$, where $\xi(r_0)=1$, matter
distribution is characterized by strong clustering; 
i.e. $\xi(r) \gg
  1$, 
about which little is known analytically and which is generally
constrained by N-body simulations where it is typically found that,
for $r<r_0$, $\xi(r) \sim r^{-\gamma}$ with $\gamma \approx 1.5$
\citep{springel}.
(ii) The  second length scale is such that $\xi(r_c)
=0$,  and it is located at $r_c \gg r_0$ \citep{pee93,glass}. 
In the range of scales $r_0 < r < r_c$, $\xi(r)$ is characterized by
positive correlations, which rapidly decay to zero when $r \rightarrow
r_c$.  This former regime can be easily related to the early universe
correlation function by a simple rescaling of amplitudes given by the
linear gravitational growth of small amplitude perturbations in an
expanding universe \citep{pee80}.  The scale $r_c$ is an imprint of
the early universe physics. It corresponds to the size of the Hubble
horizon at the time of the equality between matter and radiation and
it is fixed by the values of standard cosmological parameters being
proportional to $(\Omega h^2)^{-1}$ where $\Omega$ is the density
parameter and $h$ the normalized Hubble constant \citep{peacock}.  The
third length scale $r_{bao}$ is located on scales on the order of, but
smaller than, $r_c$. This is the real-space scale corresponding to the
baryon acoustic oscillations (BAO) at the recombination epoch.  Its
precise location depends on the matter density parameters, baryon
abundance and Hubble constant \citep{eh}.
(iii) Finally in the third range of scales, namely for $r>r_c$,
$\xi(r)$ is characterized by a negative power-law behavior,
i.e. $\xi(r) \sim - r^{-4}$ \citep{glass,book}.  Positive
and negative correlations are exactly balanced in such a way that
$\int_0^{\infty} \xi(r) d^3r =0$. This is a global condition on the
system fluctuations, which corresponds  that the matter 
distribution being super-homogeneous \citep{glass,book}
i.e. characterized by a sort of stochastic order and by fluctuations
that are depressed with respect to a purely uncorrelated distribution
of matter (i.e. white noise). This corresponds to the linear behavior
of the matter power spectrum as a function of the wave-number $k$ for
$k\rightarrow 0$ (named the Harrison-Zeldovich tail), and it
characterizes not only the LCDM model but all models of density
fluctuations in the framework of the Friedmann-Robertson-Walker metric
\citep{glass,book}.

In the new samples provided by the Sloan Digital Sky Survey Data
Release 7 (SDSS-DR7) \citep{paperdr7}, it is possible to estimate the
galaxy correlation function on scales on the order of 100 Mpc/h to
possibly determine $r_{bao}$ and $r_c$.  Some years ago,  \citet{lrg}
determined the \citet{ls} (LS) estimator of the galaxy two-point
correlation function in a preliminary luminous red galaxy (LRG) sample
of the SDSS, claiming for an overall agreement with the LCDM
prediction and for a positive detection of the scale $r_{bao}$ at
about 110 Mpc/h. More recently \citet{cabre} measured the same
estimator of the correlation function in the LRG-DR6 sample and
\citet{martinez} in the LRG-DR7 sample. They both found that the LRG
correlation is positive up to 200 Mpc/h and that the shape of the
correlation function around $r_{bao}$ is slightly different from the
one measured by \citet{lrg}.  While they claimed that the measured
correlation function was compatible with the LCDM model, they did not
discuss the fact that their detection implied that positive
correlations extend to scales larger than the model predicted
$r_c$. In addition we note that \citet{lrg,cabre,martinez} did not
discussed other estimator than the LS one.

In the present paper, we show that our results coincide very finely
with the ones of the above mentioned papers for what concern the
amplitude, shape and statistical error bars in the case of the LS
estimator in the LRG-DR7 sample.  However we measure that in the
SDSS-DR7 main galaxy (MG) sample the two-point correlation function
(LS estimator) remains positive at large separations, i.e.  for $r>
250$ Mpc/h, showing a clear systematic volume-dependent behavior and a
remarkable disagreement  with the LCDM prediction. In addition,
we find that there is a difference between the LS and the \citet{dp83}
(DP) estimator of the two-point correlation function in redshift
space. Finally we find that {\it both} estimators significantly vary
in different sky regions.  We interpret these results by studying
  the fluctuations in the sample density estimation.

The paper is organized as follows. We first define in
Sect.\ref{statmec} the estimators of the correlation function and a
simple determination of its statistical errors that we use in the data
analysis. Sect. \ref{samples} is devoted to the description of the
samples selection while in Sect.\ref{results} we present our main
results.  The discussion of the behaviors we have found and their
interpretation is presented in Sect.\ref{discussion}.  The behavior of
the two-point correlation function predicted by standard models of
galaxy formation and the comparison with the results obtained are
discussed in Sect.\ref{theo}.  Finally we draw our main conclusions in
Sect.\ref{conclusions}.


\section{Pairwise estimators} 
\label{statmec} 
In what follows we determine the two-point correlation properties by
using the LS, DP and the Hamilton (H) \citep{hamilton}
estimators. These estimators may have a number of systematic biases
when correlations are long range as we discuss in
Sect.\ref{discussion}.   Firstly, it is interesting to discuss
their properties and consider their determinations.

The LS estimator is defined as 
\be
\label{ls}
\overline{\xi_{LS} (r)} = \frac{N_r(N_r-1)}{N_d(N_d-1)}
\frac{DD(r)}{RR(r)} - 2\frac{N_r-1}{N_d} \frac{DR(r)}{RR(r)} +1 
\ee
where $DD(r)$, $RR(r)$ and $DR(r)$ are the number of data-data,
random-random and data-random pairs, and $N_r,\; N_d$ are the number
of random and data points (we use $N_r = K\cdot N_d$ with $K=3$ and we
have checked that the results do not significantly depend on $K$ as
long as this is larger than unity).

The DP estimator is defined as 
\be
\label{dp}
\overline{\xi_{DP} (r)} = \frac{N_r}{N_d-1} \frac{DD(r)}{DR(r)} -1 \;,
\ee
and the H estimator can be written as 
\be 
\overline{\xi_{H} (r)} =
\frac{N_rN_d}{(N_r-1)(N_d-1)} \frac{DD(r)RR(r)}{DR^2(r)} -1 
\;. 
\ee

 In general, a statistical estimator $X_V$ of the statistical
  quantity $X$ in a finite sample $V$, to be a valid one, must satisfy
  the following limit condition 
\[ 
\lim_{V\rightarrow \infty} X_V =\langle X \rangle \;,
\] where in
  brackets we denote the ensemble average (infinite volume limit). A
  stronger condition is that 
\[ \langle X_V \rangle = \langle X
  \rangle\;,\] 
i.e. that the ensemble average in a finite volume is equal to the
ensemble average in the infinite volume limit. If this condition is
not satisfied the estimator is said to be biased
\citep{Kerscher99,book}. One wants to understand the bias and the
variance of the various estimators and this is possible only for some
specific estimators and for distributions with simple correlation
properties (e.g. Poisson). The effect of bias, i.e. finite volume or
size effects, can be studied through the analysis of artificial
simulations with known properties; however the three estimators
defined above are all biased \citep{Kerscher99,Kerscher,cdm_theo}. It
is worth noticing that \citet{Kerscher99} showed that, in a real
galaxy sample, the three different estimators defined above use
different finite size corrections yielding to different results on
large enough scales, for small value of the correlation amplitude,
while all of them agree on smaller scales, where the amplitude of the
correlation was large enough. 

 It was shown \citep{ls} that the LS estimator has the minimal
  variance for a Poisson distribution, i.e. the variance decays as
  $1/N$ instead as $1/\sqrt{N}$ as for the DP estimator. This fact,
  however, does not mean that its variance will be any more
  controllable for a wider class of distributions with more complex
  correlation properties than Poisson's \citep{book}. 
Indeed, there is no formal proof that the DP is less accurate than the
LS for a generally correlated point distribution even though this
conclusion has been reached by, e.g., \citet{Kerscher} examining some
specific properties of estimators in Nbody simulations. They concluded
also that the H estimator is equivalent to the LS one. In
\citet{cdm_theo}, by studying finite volume effects in the estimators,
it was shown that the two estimators LS and the H are indeed
indistinguishable, but that they are almost equivalent to the DP when
the underlying distribution is positively correlated.

Among the various ways to compute statistical errors
\citep{cdm_theo} we use the jack-knife (JK) estimate  whose
variance is \citep{scranton}
\be
\label{jackerrors}
\sigma^2_{Jack} (r)= \sum_{i'=1}^{N} \frac{DR_{i'}(r)} {DR(r)} \left(
\overline{\xi^i_{LS} (r)}  -  \overline{\xi_{LS} (r)}  \right)^2 
\ee 
where the index $i$ is used to signify that the value of the correlation
function  is computed each time in all the
$N$ sub-samples of a given samples but one (the $i^{th}$).


\section{The samples} 
\label{samples}

We have constructed several sub-samples of the main-galaxy (MG) and
the luminous-red-galaxy (LRG) samples of the spectroscopic catalog
SDSS-DR7. Concerning the latter we have constrained the flags
indicating the type of object to select only the galaxies from the MG
sample.  We then consider galaxies in the redshift range $10^{-4} \leq
z \leq 0.3$ with redshift confidence $z_{conf} \ge 0.35$ and with
flags indicating no significant redshift determination errors.  In
addition we apply the apparent magnitude filtering condition $r <
17.77$ \citep{strauss2002}.

The angular region we consider is limited, in the SDSS internal
angular coordinates, by $-33.5^{\circ} \le \eta \le 36.0^\circ$ and
$-48.0^\circ \le \lambda \le 51.5^\circ$: the resulting solid angle is
$\Omega=1.85$ steradians.  We do not use corrections for the redshift
completeness mask or for fiber collision effects. Fiber collisions in
general do not present a problem for measurements of large scale
galaxy correlations \citep{strauss2002}.  Completeness varies most
near the current survey edges which are excluded in our samples.
 The completeness mask takes into account that the fraction of
 observed galaxies is not the same in all the fields, because of both
 fiber collision effects and small variation in limiting
 magnitude. One can, under certain assumption, take into account the
 completeness mask information in the statistical analysis.  Otherwise
 it is possible to make tests by varying the limits in apparent
 magnitude and study the stability of the results obtained. We have
 applied this second possibility and we did not find sensible
 variations in the measured statistical properties when $r<17.5$
 \citep{sdss_aea}. This conclusion is confirmed by the fact that our
 results for the LRG sample agree with those of
 \citet{lrg,cabre,martinez} and for the MG sample with those of
 \citet{zehavi2,zehavi3}, who have explicitly taken into account the
 completeness mask of the survey in their analysis. As noticed by \citet{cabre} 
 the completeness  mask
 could be the main source of systematic effects on small scale only,
 while we are interested on the correlation function on relatively
 large separations.

To construct volume-limited (VL) samples (see
Tab.~\ref{tbl_VLSamplesProperties1}) we computed the metric distances
using the standard cosmological parameters, i.e., $\Omega_M=0.3$ and
$\Omega_\Lambda=0.7$ with $H_0=100 h$ km/sec/Mpc. We computed absolute
magnitudes using Petrosian apparent magnitudes in the $r$ filter
corrected for Galactic absorption.  

We checked that the main results in the MG sample we got do not depend
on K-corrections and/or evolutionary corrections as those used by
\citet{blanton2003}. In this paper we use standard K-correction from
the VAGC data \footnote{{\tt http://sdss.physics.nyu.edu/vagc/}} (see
  discussion in \citet{sdss_aea} for more details).

\begin{table}
\begin{center}
\begin{tabular}{|c|c|c|c|c|c|}
  \hline
  VL sample & $R_{min}$ & $R_{max}$ & $M_{min}$ 
& $M_{max}$ & $N$ \\
  \hline 
  VL1    & 50  & 200 & -18.9 & -21.1  & 72037 \\
  VL2    & 150 & 500 & -21.1 & -22.4  & 69999 \\
  VL3    & 200 & 600 & -21.5 & -22.7  & 42357\\
  VL4    & 70  & 450 & -20.8 & -21.8  & 93821\\  
  LRG   &  570 & 1035 & -20.5 & -22.5 & 53066\\
\hline
\end{tabular}
\end{center}
\caption{Properties of the SDSS-DR7 VL samples: $R_{min}$,
  $R_{max}$ (in Mpc/h) are the metric distance limits;
  ${M_{min}, \,M_{max}}$  the absolute
  magnitude limits in the $r$ filter; $N$ is the number of
  galaxies.}
\label{tbl_VLSamplesProperties1}
\end{table}

Concerning the LRG we have selected all the objects that have
classification ``galaxy'' and which belong to the ``Cut I'' subset of
the Galaxy Red objects with the same redshift quality
criteria as for main galaxies.  As this is only roughly VL
 sample we have applied cuts in absolute magnitude $M$ 
and distance $R$ to obtain a rectangular area in the $M-R$ diagram.
In addition because evolutionary effects are small for LRG galaxies
\citep{lrg} we have not applied further corrections to these data.
Given that we have selected a truly VL sample, we did not apply 
a further redshift dependent weighting to the data.

The sub-samples used to measure the JK errors are made by
dividing the survey angular region we considered into 30 sub-fields,
each of area $\sim$ 200 deg$^2$. In this way there are some thousands
galaxies in each sub-sample.


\section{Results} 
\label{results}

\begin{figure}
\begin{center}
\includegraphics*[angle=0, width=0.5\textwidth]{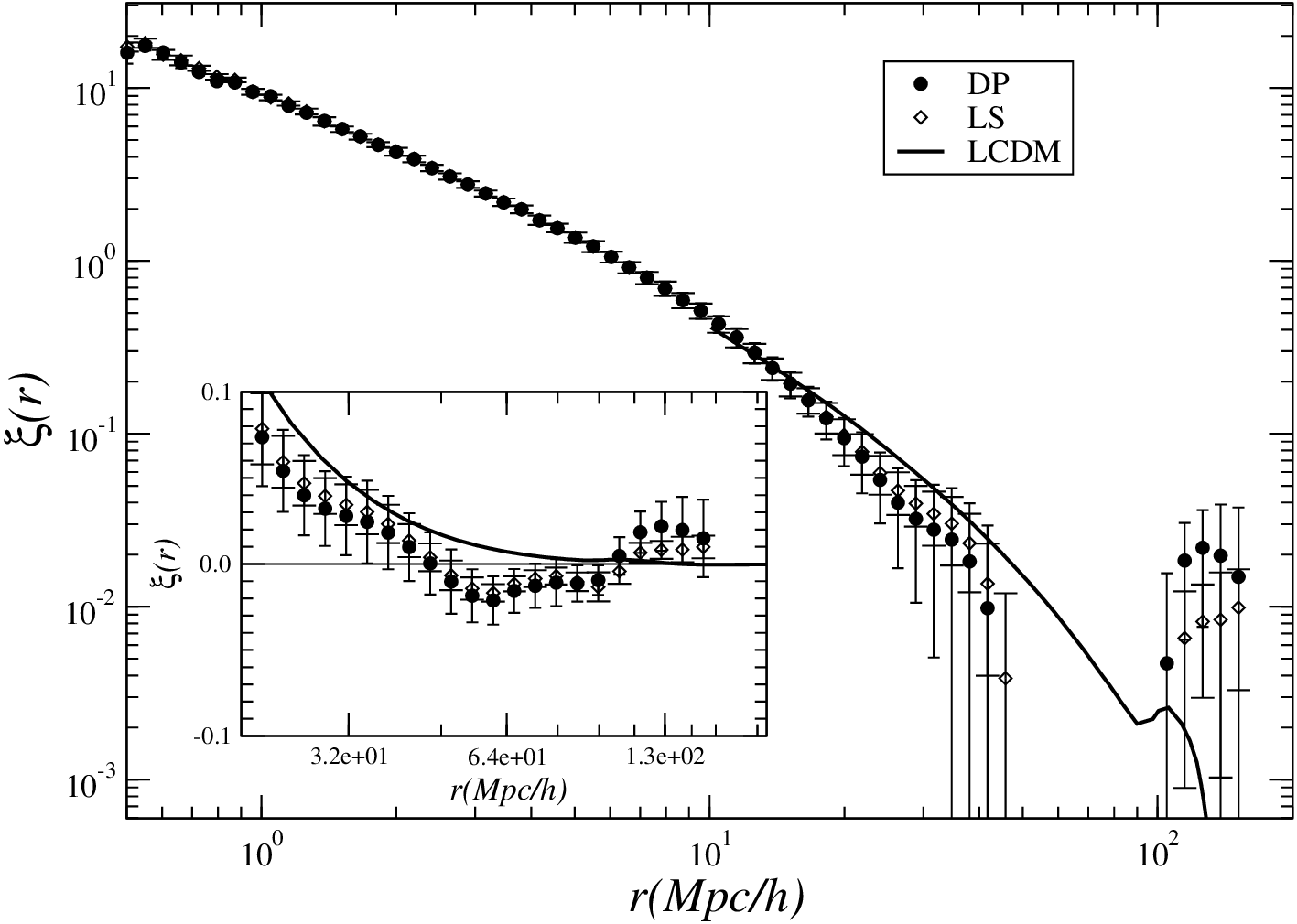}
\end{center}
\caption{Correlation function in the VL1 sample: both the LS and the
  DP estimators are reported.  The solid line gives prediction of the
  LCDM with $\Omega_m h^2= 0.12$ (from \citet{lrg}) linearly rescaled,
  according to the simplest biasing scheme \citep{kaiser}, to fit the
  amplitude on 10 Mpc/h. In the insert panel we show the same behavior
  but in a log-linear scale.}
\label{fig1}
\end{figure}

\begin{figure}
\begin{center}
\includegraphics*[angle=0, width=0.5\textwidth]{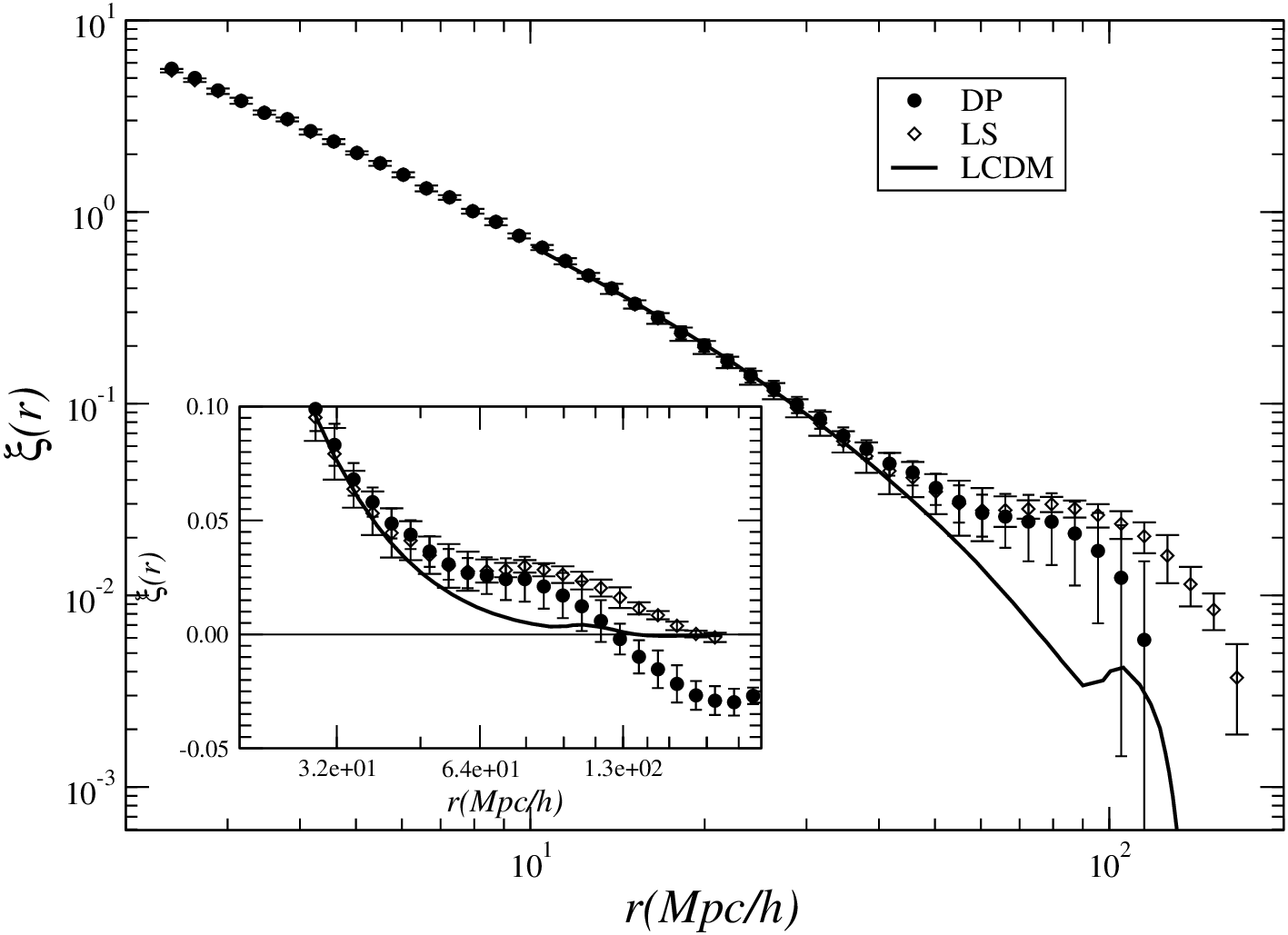}
\end{center}
\caption{The same of Fig.\ref{fig1} but for the VL2 sample.}
\label{fig2}
\end{figure}

\begin{figure}
\begin{center}
\includegraphics*[angle=0, width=0.5\textwidth]{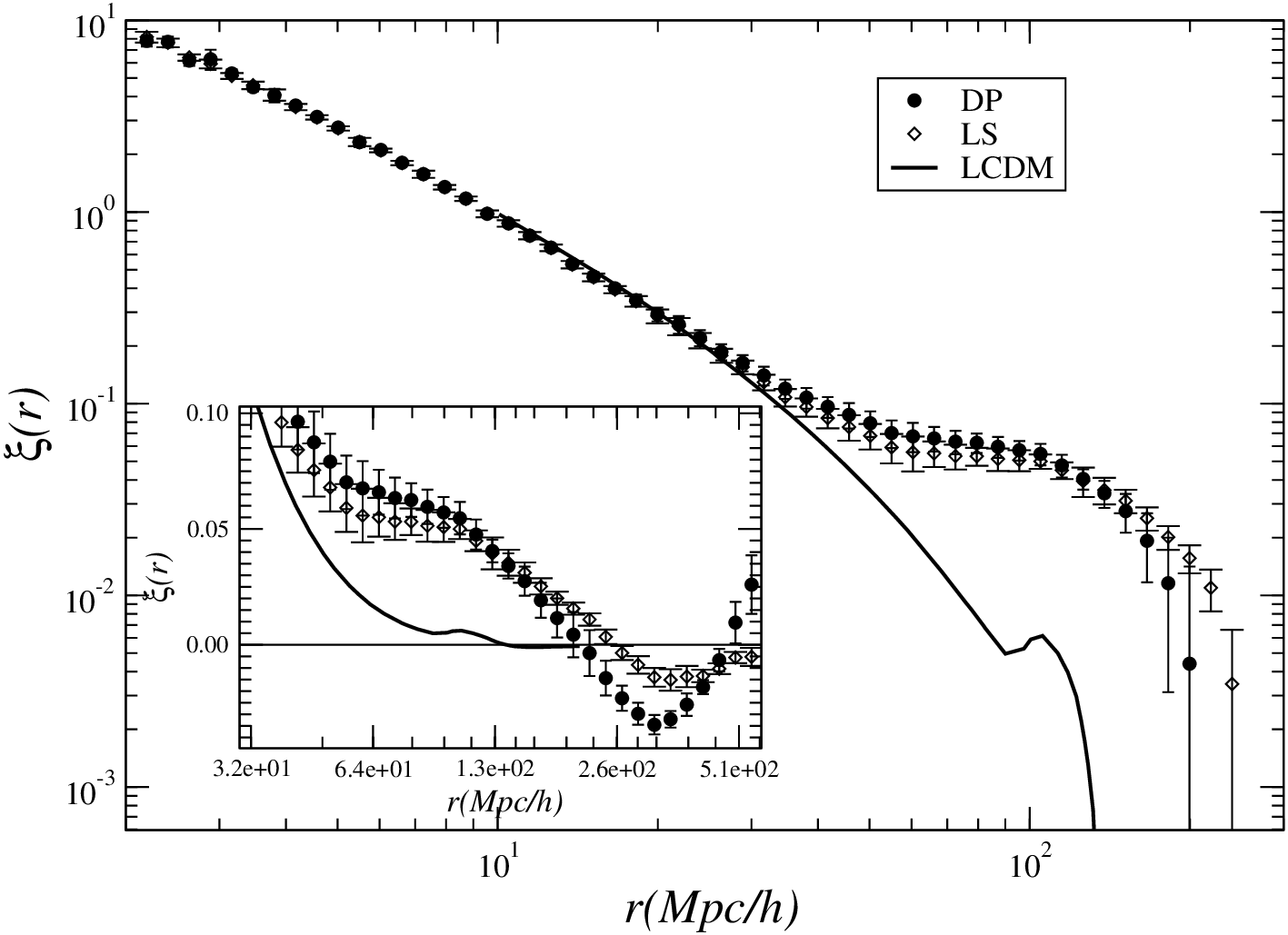}
\end{center}
\caption{The same of Fig.\ref{fig1} but for the VL3 sample.}
\label{fig3}
\end{figure}

\begin{figure}
\begin{center}
\includegraphics*[angle=0, width=0.5\textwidth]{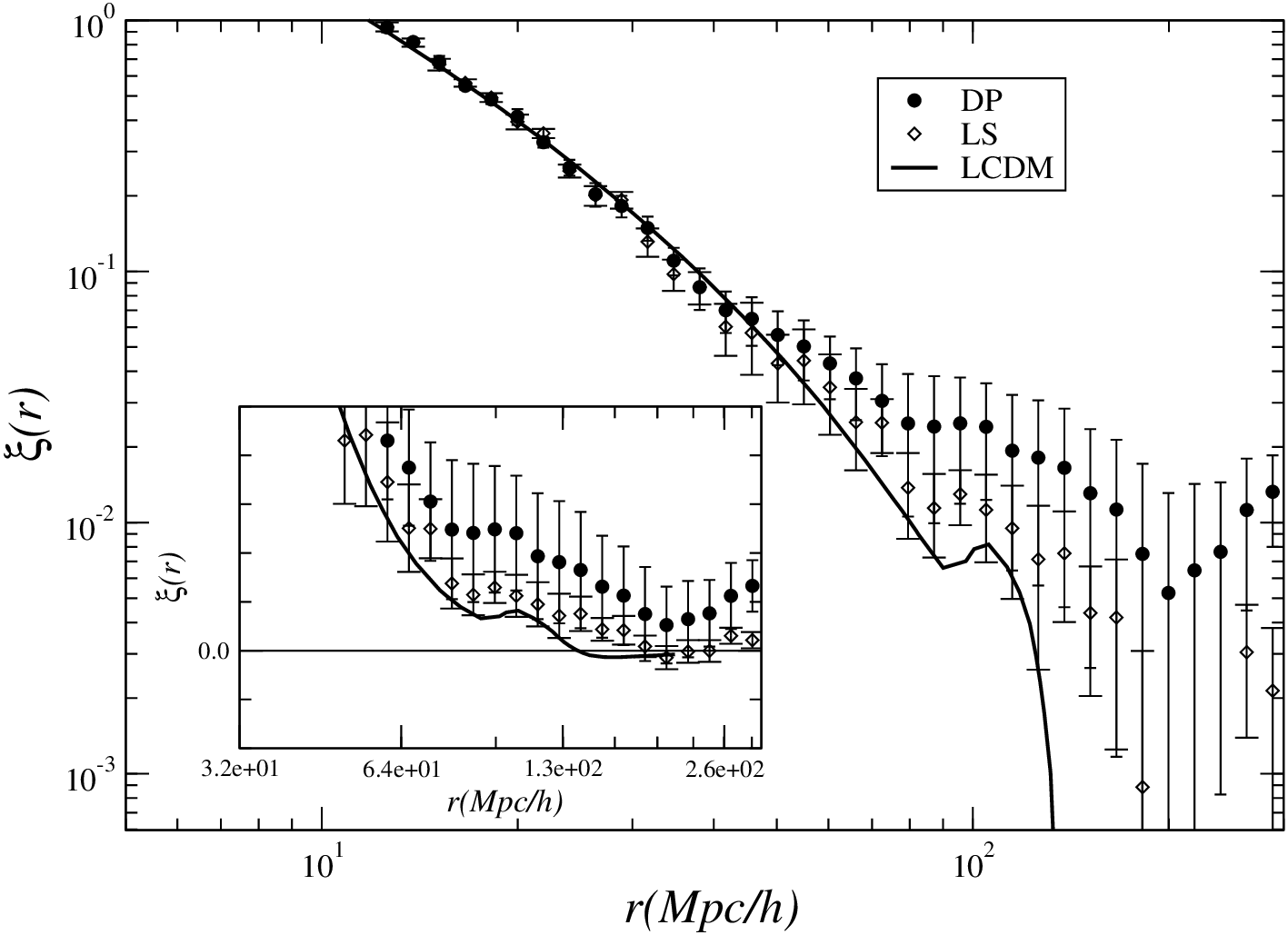}
\end{center}
\caption{The same of Fig.\ref{fig1} but for the LRG sample.}
\label{fig4}
\end{figure}

\begin{figure}
\begin{center}
\includegraphics*[angle=0, width=0.5\textwidth]{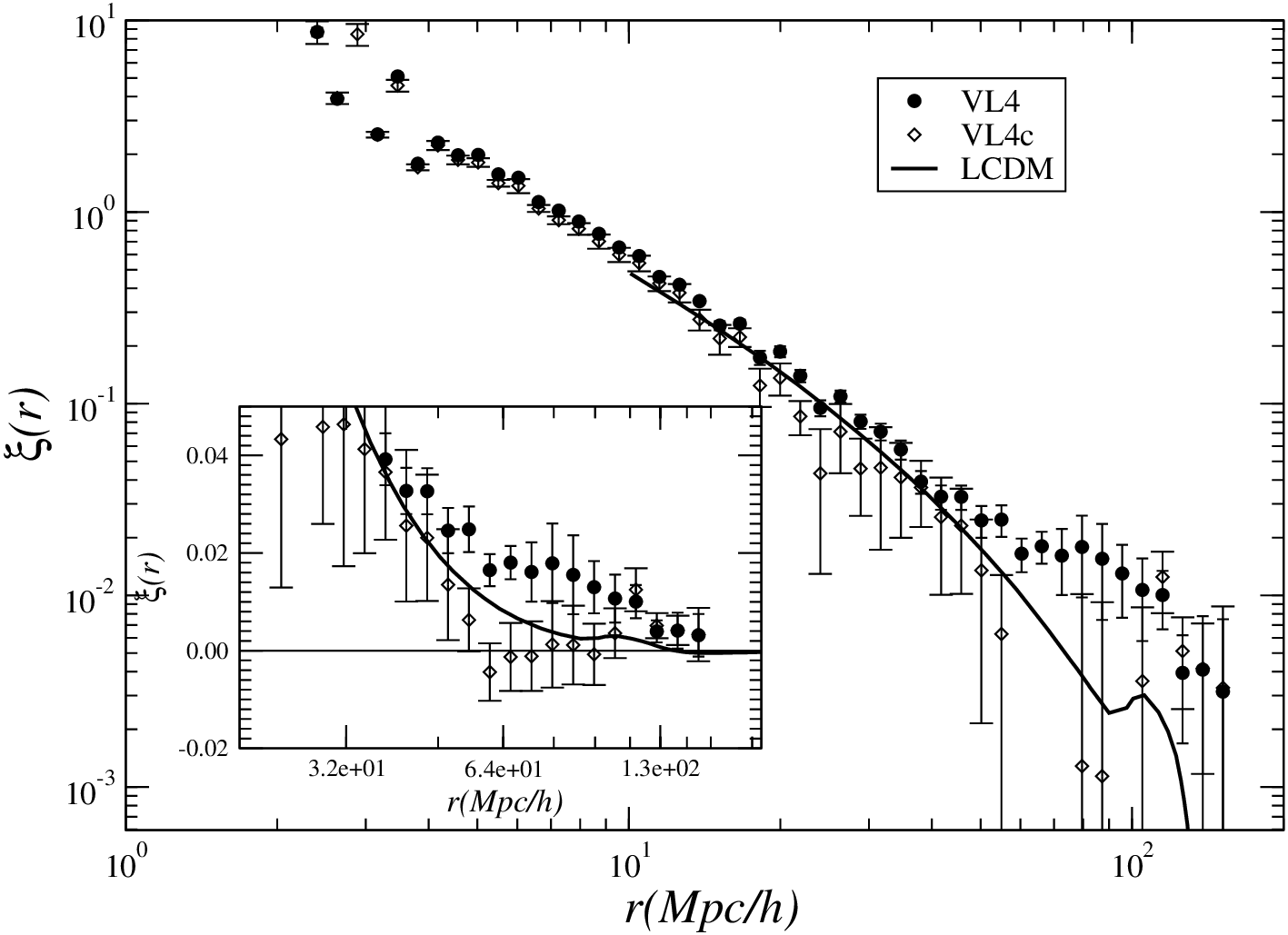}
\end{center}
\caption{Correlation function in the whole sample VL4 and in a sub-sample of it (VL4c)  
limited at $R_{max} =250$ Mpc/h. Jack-knife errors are
shown in both cases. } 
\label{figvl4}
\end{figure}

\begin{figure}
\begin{center}
\includegraphics*[angle=0, width=0.5\textwidth]{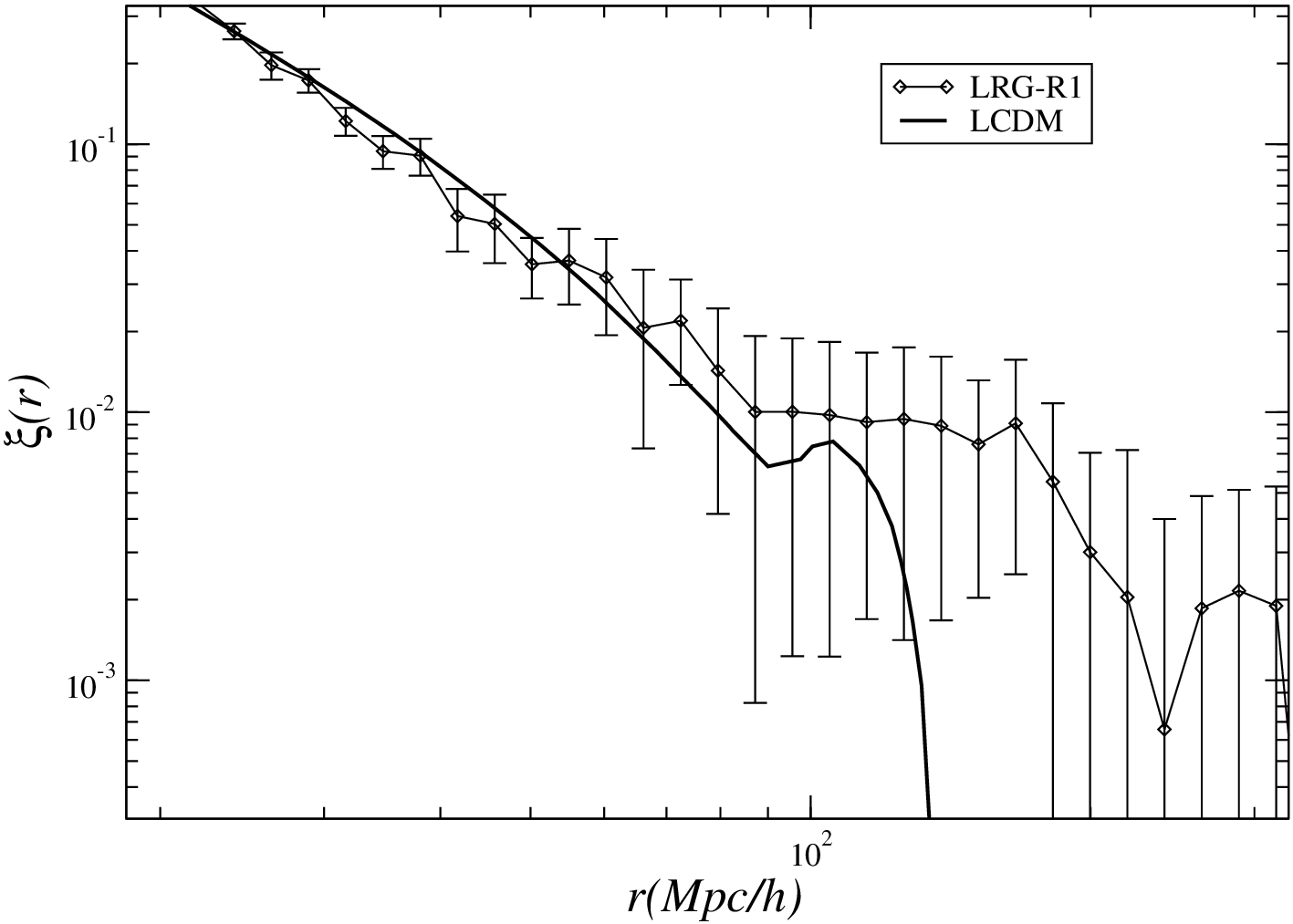}
\end{center}
\caption{ Correlation function measured through the LS estimator (with
  jack-knife errors) in the LRG sub-sample (R1) which is limited by
  $-33.5^{\circ} \le \eta \le 36.0^\circ$ and $-48.0^\circ \le \lambda
  \le 0^\circ$, i.e. with solid angle $\Omega =0.9$ steradians. The
  solid line is the LCDM prediction.  }
\label{fig5}
\end{figure}

\begin{figure}
\begin{center}
\includegraphics*[angle=0, width=0.5\textwidth]{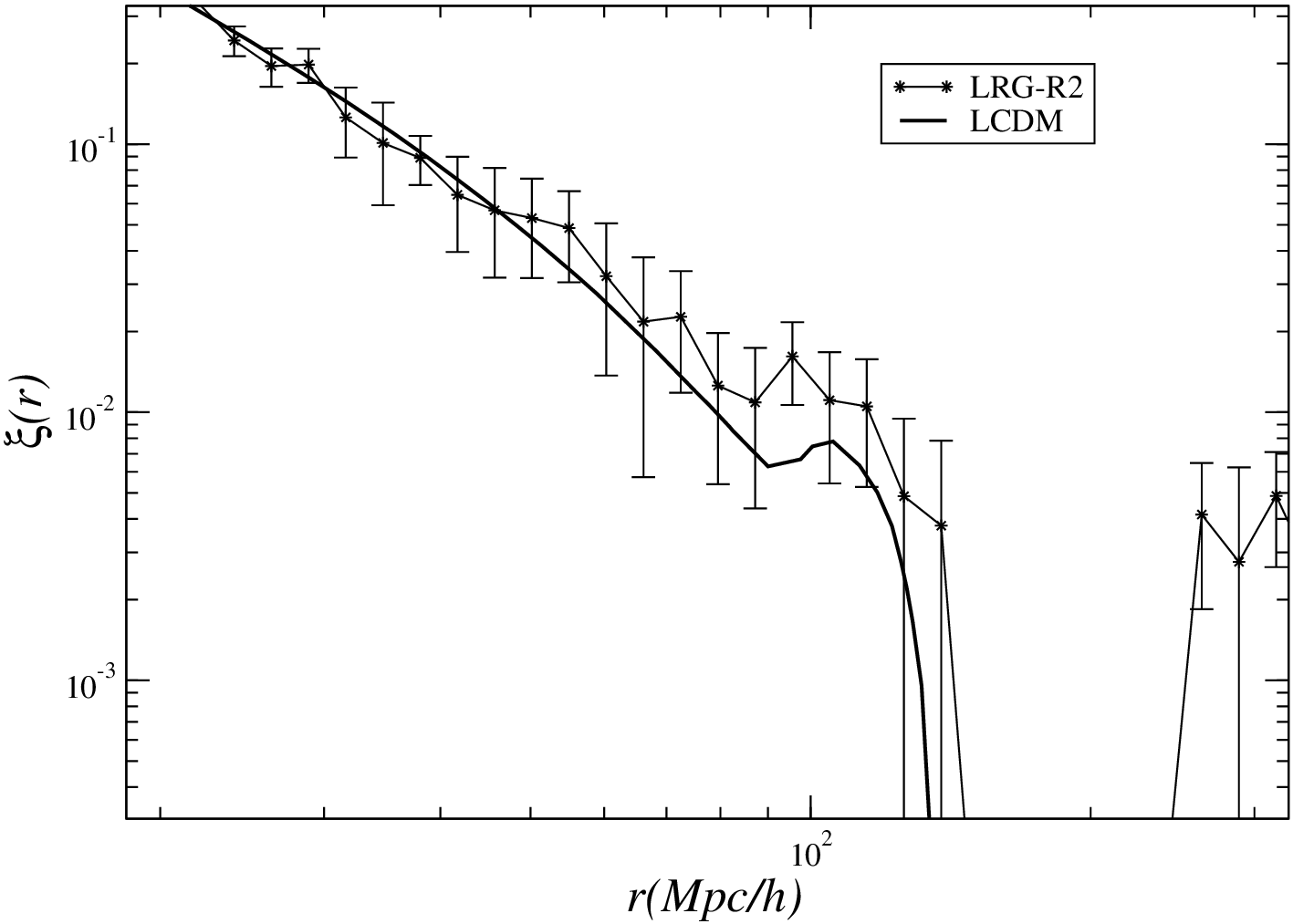}
\end{center}
\caption{ The same of Fig.\ref{fig5} but for the R2 angular region,
  which is limited by $-33.5^{\circ} \le \eta \le 36.0^\circ$ and
  $0^\circ \le \lambda \le 48^\circ$, i.e. with solid angle $\Omega
  =0.9$ steradians.  }
\label{fig6}
\end{figure}

We find, in agreement with \citet{zehavi1,zehavi2,lrg} in previous
data releases of the SDSS, that the redshift-space correlation
function in different samples shows a different amplitude but similar
shape on small scales (see Figs.\ref{fig1}-\ref{fig4}). This is
usually ascribed to the (physical) effect of selection, that brighter
galaxies exhibit a larger clustering amplitude
\citep{zehavi1,zehavi2,norberg2}. However this is not the only change:
the larger the correlation function amplitude the more extended is the
range of scales where there are detectable (i.e signal larger than
JK) positive correlations.  Indeed, in the MG samples
the transition scale from positive to negative correlations occurs at
a scale that grows roughly in proportion to the sample size and in the 
deepest samples this is located on rather larger  scales, i.e.  $r>250$ Mpc/h.
However in the VL1 sample we find $r_c\approx 50$ Mpc/h, i.e. less than 
the half of the LCDM prediction.

To show that finite-volume effects are important on large separations, 
we consider a single sample (VL4)
and we cut it at different scales $R_{max}$; in addition we consider
an angular cut of the LRG sample for which the depth is fixed but the
volume is lowered.  In the latter case the whole angular region of
$\sim 6000$ deg$^2$ is cut into two non-overlapping sky region, each
of area $\sim 3000$ deg$^2$, i.e only $\sim 20\%$ smaller than the
sample considered by \citet{lrg}.
As one may notice from Figs.\ref{figvl4}-\ref{fig6}, there is a clear
volume dependence of the two-point correlation function on large
scales.   In
particular, in the R1 sub-sample there is an evident difference between
the data and the LCDM prediction.
In addition, we note that almost in all cases the DP and LS
estimator on large enough scales show a difference which can be larger
than statistical error bars.

It is worth noticing that our result for the LS estimator of the
correlation function in the LRG sample finely agrees with the
determination of \citet{martinez}, although these authors have used a
slightly different technique to take into account the survey
completeness mask, as we commented above (see Fig.\ref{fig7}).
\begin{figure}
\begin{center}
\includegraphics*[angle=0, width=0.5\textwidth]{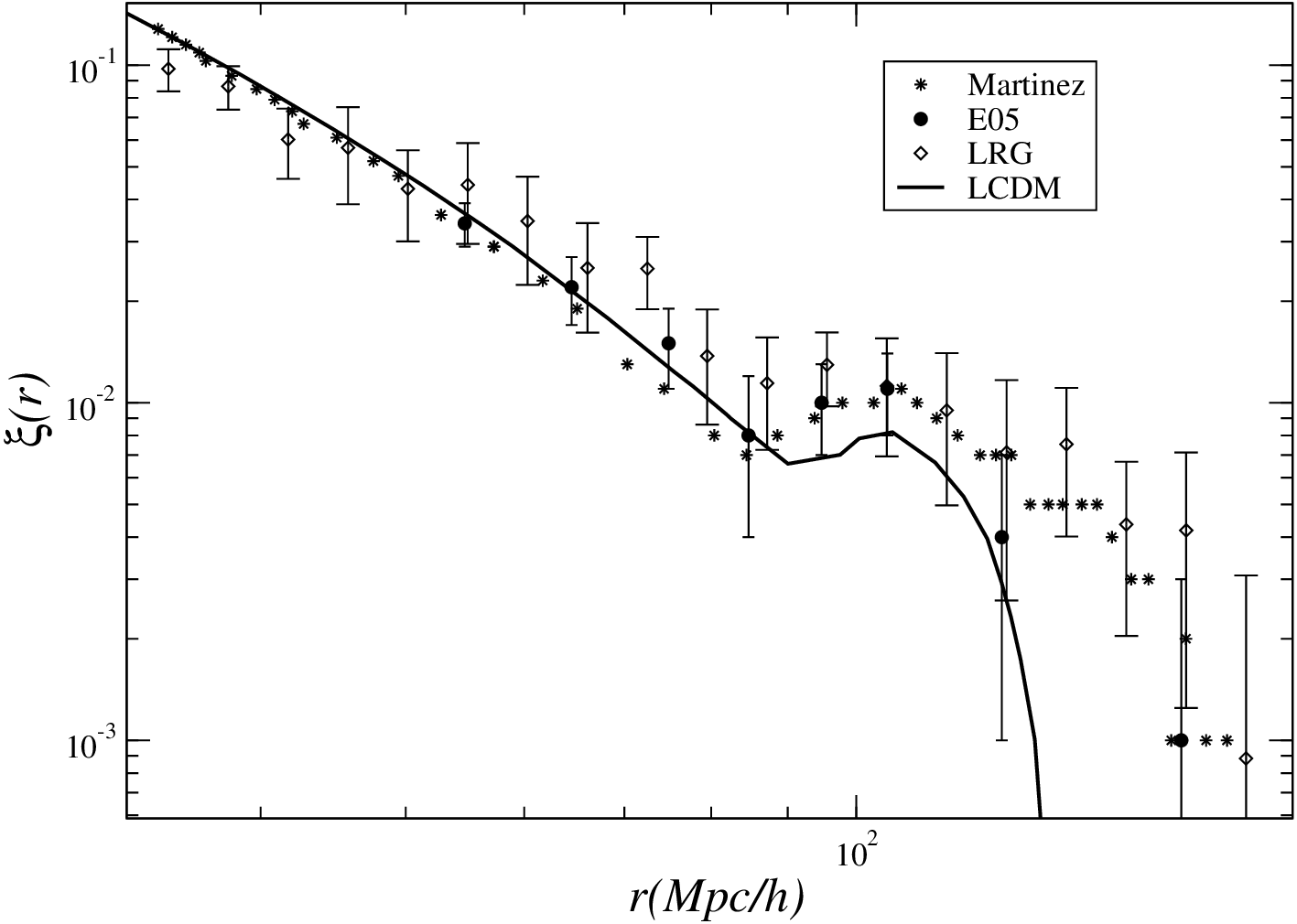}
\end{center}
\caption{Determination of the correlation function for the LRG sample
  with the LS estimator (LRG), compared with the \citet{lrg} (E05) and the
  \citet{martinez} (Martinez) determinations. The solid line is the
  LCDM prediction.}
\label{fig7}
\end{figure}
The LS estimator for the LRG sample is also very similar to the
determination made by \citet{lrg}, although the signal is larger then
the statistical error bars and positive up to 200 Mpc/h, as it was
found also by \citet{martinez} in the same sample we considered. A
similar trend was also seen in the analysis by \citet{cabre}. 
In addition our result for the MG sample nicely agree with the
determination of \citet{zehavi3}, although they did limit their
analysis to smaller scales than the ones considered in our analysis.

We note that \citet{lrg} stated that the MG sample does not have an
enough large volume to measure the correlation function on 100 Mpc/h
scales, without giving a clear quantitative argument of why statistical or
systematic errors should prevent one to measure the correlation
function on those scales. Indeed, we find that the signal to noise
ratio, when JK error estimations are used, is larger than
unity even on scales larger than 150 Mpc/h. In this respect one may
ask whether statistical errors computed in this way are meaningful.  

In addition we note that \citet{martinez} also found that the
correlation function becomes negative on scales of the order $50$
Mpc/h in a 2dFGRS sample, without however commenting on this
fact. Actually they even claimed that $r_{bao}$ is detectable when the
correlation function is negative, without discussing that this is not
what one expects in the context of the LCDM model where the zero point
of the correlation function must be a single scale for any type of
objects (see below).

Finally we find that, as discussed in Sect.2, the LS and the H
  estimators of the correlation function are almost indistinguishable:
  this is shown in Fig.\ref{fig_ham} where we plot the behavior of the
  ratio $\overline{\xi_{LS} (r)}/\overline{\xi_{H} (r)}$ as a function
  of separation. This remains smaller than $\sim 5\%$ on all the
  relevant scales.

Finally to check the number of points $N_r$ used in the random
  sample do not alter the estimation we have increased $N_r$ up to ten
  times the number of $N_d$ without detecting any sensible change. We
  conclude that the difference between the DP and LS estimator lies in
  the bias (finite-volume effect) intrinsic to the different ways
  these estimators take into account boundary conditions.

\begin{figure}
\begin{center}
\includegraphics*[angle=0, width=0.5\textwidth]{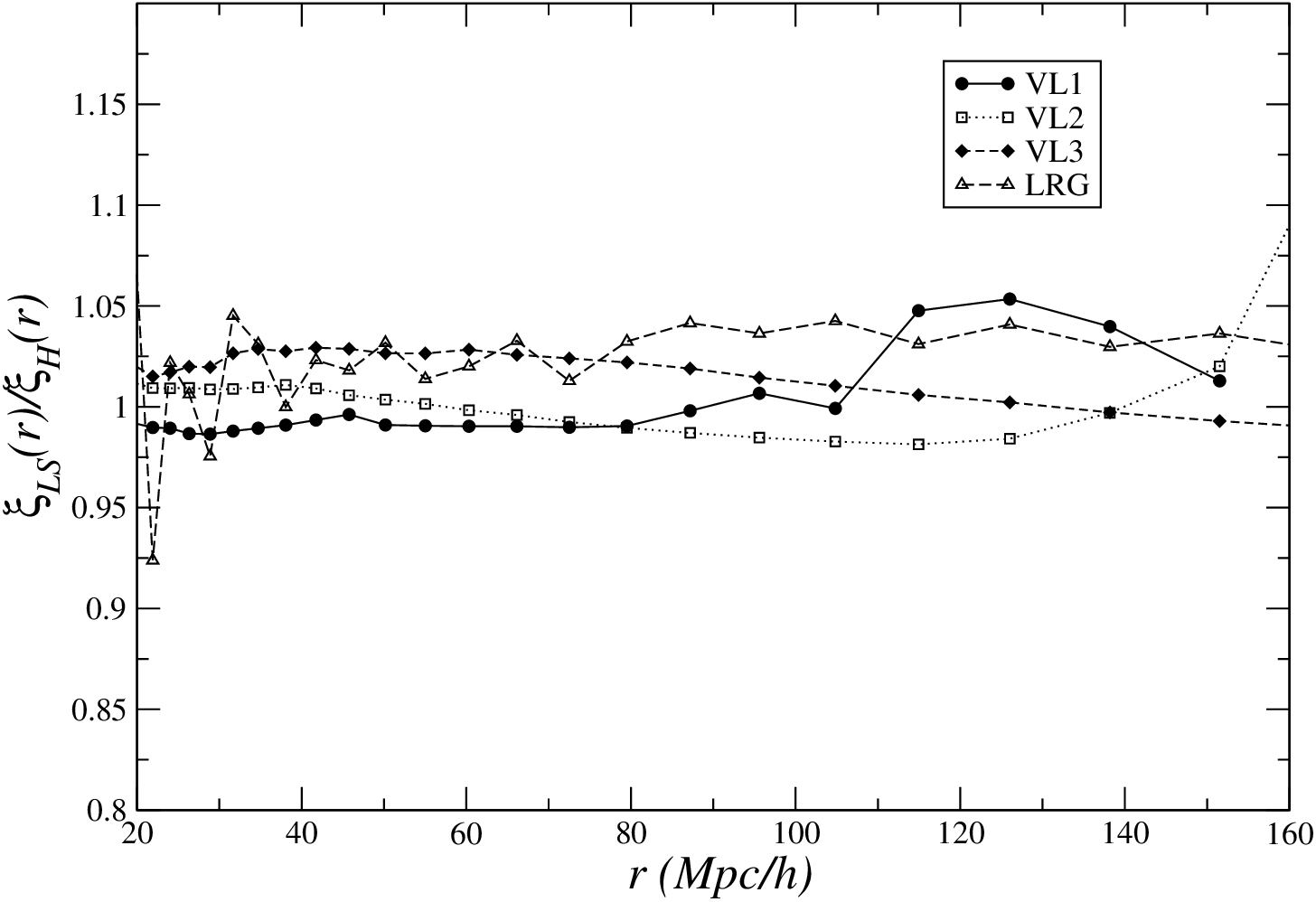}
\end{center}
\caption{Behavior of the ratio
    $\overline{\xi_{LS}(r)}/\overline{\xi_{H} (r)}$ as a function of
    separation in the different MG and LRG samples. }
\label{fig_ham}
\end{figure}
%


\section{Fluctuations and  volume-dependent systematic effects} 
\label{discussion} 

In theoretical models, the matter density field is uniform on large
scales and the average mass density $\langle n \rangle$ is provided by
an average over an ensemble of realizations of a given stochastic
process. In a finite sample of volume $V$, the average density
$\overline{n}$ can be estimated in some way. In the limit in which the
sample volume is infinite and in the process is ergodic \citep{book}
then
%
$\lim_{V \rightarrow \infty}\overline{n} =
\langle n \rangle $
%
because in this limit the relative variance goes
to zero if the distribution is uniform on large scales, i.e.  
\be
\label{eq:sigma} 
\lim_{V \rightarrow \infty} \overline{\sigma^2(V)} = \lim_{V \rightarrow \infty} 
\frac{ \overline{ \Delta N(V)^2 } }{\overline{ N(V)}^2 }  
=0 
\ee 
where $N(V)$ is the mass in a volume $V$.  In a finite volume
$\overline{\sigma^2(V)}$ is finite and therefore in any finite volume
$\overline{n} \ne \langle n \rangle$. In general for a uniform
stochastic point process, in the ensemble average sense 
the relative mass variance can be written as 
\be
\label{eq:sigma2} 
\sigma^2(V)= \frac{1}{V^2} \int_V \int_V \xi(\vec{r}_1-\vec{r}_2)
d^3r_1 d^3r_2 + \frac{1}{\langle N(V) \rangle} 
\ee 
where $\xi(r)$ is the ensemble average two-point correlation function.
In the r.h.s. of Eq.\ref{eq:sigma} there is the sum of the
contribution to the variance due to correlation and due to Poisson
noise, the former being always present in a point distribution.

Thus in a finite sample any determination of the average density
$\overline{n}$ has an intrinsic error $\overline{\sigma(V)}$. Given
that the two-point correlation determines the amplitude of
correlations with respect to the sample density, it is natural to ask
which is the error introduced in the estimation of the correlation
function by the  uncertainty on the value of the sample density. A
second question is which kind of statistical estimation of the
correlation function errors in a finite sample is representative of
the errors induced by the average density uncertainty.

\subsection{Fluctuations in the determination of the sample density}

The two-point correlation function is defined as 
\be
\label{xi1} 
\xi(r) = \frac{ \langle n(r)n(0) \rangle} {\langle n \rangle^2} -1 =
  \frac{\langle n_p(r) \rangle } {\langle n \rangle} -1 
 \ee
where 
\[
\langle{n_p(r)}\rangle = \frac{\langle{ n(r)n(0) \rangle}}{\langle n \rangle}
\] 
is the conditional density. Because of the  definition in Eq.\ref{xi1}, any
 estimator of $\xi(r)$ can be written as
\be
\label{xi2} 
\overline{\xi(r)} =  \frac{\overline{n_p(r)}} {\overline{n_s}} -1 
\ee
where $\overline{n_p(r)}$ is the sample estimation of the conditional
density and $\overline{n_s}$ is the sample estimation of density.
Note that, in general, to measure the conditional density, one
performs an average over all points in the sample \citep{book}.
On the other hand the estimation of the sample average does not
involve the average operation. For instance one can simply determine
the sample density to be $\overline{n_s} = \frac{N}{V}$ where $V$ is
the sample volume and $N$ is the number of objects in it.

In addition it is worth noticing that the pair-wise estimators
introduced in Sect.\ref{statmec}, necessarily use a similar strategy,
as in order to the measure the average of the sample density one would
need many samples of size $V$.  Thus, the determination of the
two-point correlation function requires the estimation of an average
quantity and of a non-average quantity. The former can introduce
volume-dependent systematic effects in a non-trivial way.

Suppose that a certain estimator $\overline{\xi_{1}(r)}$ of the two-point
correlation function uses the sample estimation $\overline{n_1}$ while
another estimator $\overline{\xi_{2}(r)}$ uses $\overline{n_2}$: the
difference between $\overline{n_1}$ and $\overline{n_2}$ is not due to
the fact that the samples are different, rather that the different
estimators use different boundary conditions to measure the two-point
correlation function, i.e.  different ways of normalizing the
data-data pairs to the data-random and random-random pairs.
Thus they are subject to a different bias \citep{Kerscher99}.
  Alternatively one can think to measure the same estimator but into
  two different samples of same geometry and volume, in which the
  sample density takes a slightly different value.

We show now that the different values the sample density may
result in a different measurement of the large scales behavior of the
correlation function. To this aim, let us assume that there is a small
difference between the value of the sample density used by the
estimator 1 and the estimator 2, so that we can write
\be 
\label{err} 
\overline{n_2} = \overline{n_1} (1+\delta) 
\ee 
with $\delta \ll 1$.  Le us also suppose that the two estimators
measure the exactly same conditional density $\overline{n_p(r)}$.
This is a simplifying but reasonable assumption as the conditional
density is averaged over many points placed in different 
parts of the sample volume. In these conditions we may write that 
\be
\overline{\xi_{1,2}(r)} = \frac{\overline{n_p(r)}}
         {\overline{n_{1,2}}} -1 
\ee 
and thus from Eqs.\ref{xi1}-\ref{err} we get
\be
\label{xi3} 
\overline{\xi_2(r)} = \frac{1+\overline{\xi_1(r)}}{1+\delta} -1
\approx \overline{\xi_1(r)} -\delta 
\;, 
\ee
which makes explicit that a different determination of the sample
density results in a variation of the estimated two-point correlation
function.

As an illustrative example, we can take as $\overline{\xi_1(r)}$ the
LS estimator for the LRG sample. We find that, for $\delta= - 0.006$,
$\overline{\xi_2(r)}$ in Eq.\ref{xi3} almost perfectly agrees with the
DP estimator in the same sample (see Fig.\ref{xilsdp}).  It is thus
clear than a small  uncertainty in the value of the sample
average (in this case 0.6 \%) can affect the large scale behavior of
the correlation function in the range of scales and of amplitudes of
interest, i.e. around 100 Mpc/h in the LRG sample.  Therefore, we have
to determine what is the error on the estimation of the sample density
and then we have to clarify how this changes the large scale behavior
of the correlation function. {\it Is the above estimation of 0.6\%
  representative of the true  uncertainty on the large scale
  average density ?}

Simply stated, the problem is the following: in order to measure the
BAO we need to have an error of about $10^{-3}$ on the estimator of
the correlation function. Indeed, for the LRG case, the correlation
function on 100 Mpc/h has an amplitude of about $10^{-2}$ while the
feature corresponding to the BAO (a slight local increase followed by
a decrease) corresponds to a local variation of about $10^{-3}$ in the
correlation function amplitude.

By errors propagation, we find from Eq.\ref{xi2} that
\be
\label{xie1} 
\overline{\delta \xi(r)} \simeq \frac{\delta \overline{n_p(r)}}
         {\overline{n_s} } + \frac{\overline{n_p(r)}} {\overline{n_s}}
         \overline{\sigma} \;.  \ee
We neglect again the statistical error on the determination of the
conditional density on scales smaller than the sample, i.e. the first
term in the r.h.s. of Eq.\ref{xie1}. As discussed above, this
approximation is reasonable in view of the fact that the conditional
density is determined by making an average over many points.  Then by
using again Eq.\ref{xi2} we can rewrite the previous equation as
\be
\label{xie2} 
\overline{\delta \xi(r)} \simeq ( \overline{\xi(r)} +1 )  \overline{\sigma}
\approx \overline{\sigma}
\ee
where we used that $\overline{\xi(r)} \ll 1$ as this is the regime in
which we are interested in. From Eq.\ref{xie2}, it follows that the
error on the correlation function estimation is of the same order of
the error in the estimation of the sample density. Therefore the
question is whether we really know the sample density with an error of
the order of $10^{-3}$.
\begin{figure}
\begin{center}
\includegraphics*[angle=0, width=0.5\textwidth]{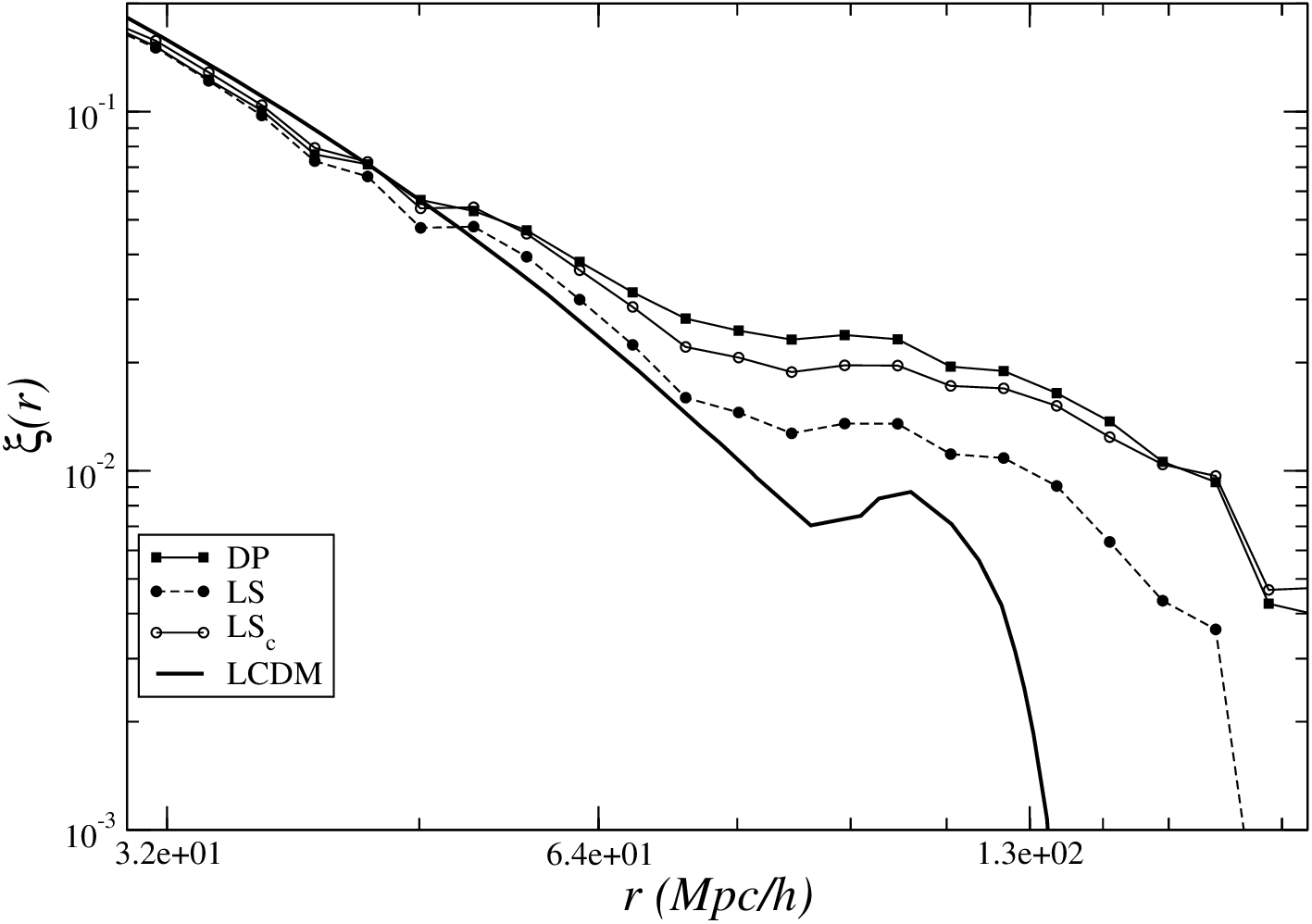}
\end{center}
\caption{ By taking the LS estimator in the LRG sample for
  $\overline{\xi_1(r)}$ in Eq.\ref{xi3} we find that for $\delta= -
  0.006$ the quantity $\overline{\xi_2(r)}$ (labeled as LS$\_c$)
  almost perfectly agrees with the DP estimator in the same sample.}
\label{xilsdp}
\end{figure}

The typical fluctuation on the density
estimation in a given sample, on scales of the order of the sample
size, is $ \overline{\sigma}$. The problem is to constrain
$\overline{\sigma}$ from the data.  As mentioned above it is not
possible to make an average over many samples of volume $V$, as we
have a single one, and thus we can determine the fluctuation only
inside the sample itself by considering several sub-samples of it.

We have estimated $\overline{\sigma}$ on the relevant scales as
follows. We divide the sample into $N$ independent (non-overlapping)
angular fields and then we determine the number of galaxies in the
each field. We then compute the average $\overline{N}$ and the
variance $\overline{\Sigma^2}$ and thus the standard deviation as
\be
\label{xie3}
\overline{\sigma}=\frac{\sqrt{\overline{\Sigma^2}}}{\overline{N}} \;.
\ee 
As there is an arbitrariness in the choice of the number of fields $N$
we let it to vary between a few, for which we have more than $10^4$
objects in each field, to some tens, to have a least several hundreds
of galaxies in each field.

From Fig.\ref{flucts} we may note that in the LRG sample, the typical
fluctuation is about $8 \%$ for about any value of $N$ and that this
is much larger than Poisson noise, i.e. almost a factor 100 larger 
than the error needed to measure the correlation function
with a precision of the order of $10^{-3}$ !

\begin{figure}
\begin{center}
\includegraphics*[angle=0, width=0.5\textwidth]{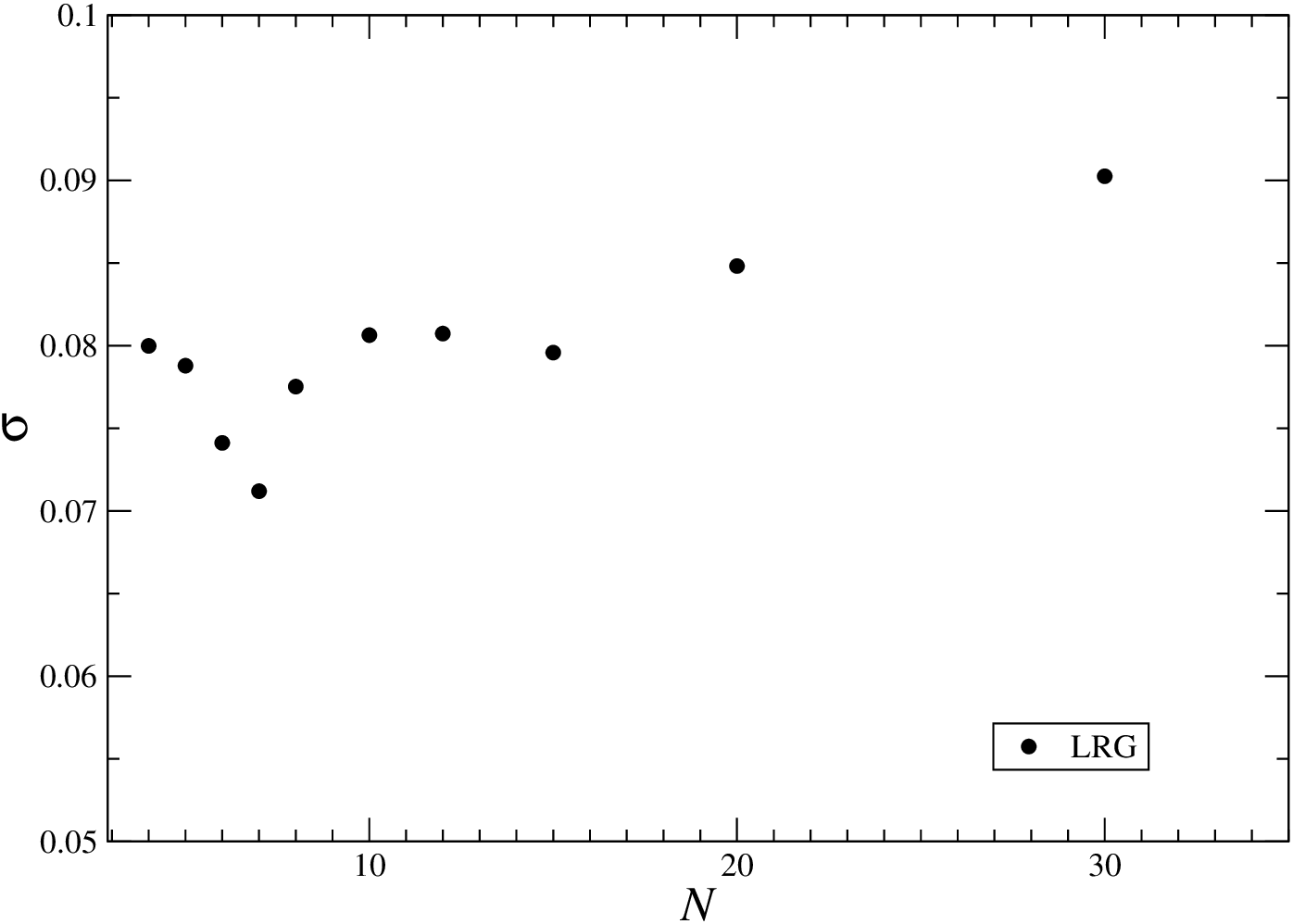}
\end{center}
\caption{The typical fluctuation $\overline{\sigma}$ in the LRG sample
  average density is about $8\%$ for about any value of $N$ in the
  range $4,30$ and it is much larger than Poisson noise.}
\label{flucts}
\end{figure}

Note that this value of the typical fluctuation is in agreement with
that obtained in a smaller LRG sample by \citet{hogg}.  For the MG
samples we find that $\overline{\sigma}$ has about the same amplitude
as for the LRG case (see Fig.\ref{flucts2}).  Thus given that
$\overline{\sigma} \ltapprox 0.1$ we conclude that we can get in these
samples a statistically significant estimation of the correlation
function only for $|\xi(r)|\gtapprox |\delta \xi(r)| \approx
\overline{\sigma} \approx 0.1$ and thus any claim about smaller
amplitude is biased by overall volume-dependent systematic
effects. This implies that, for the LRG sample, our estimation is
statistically significant for $r \ltapprox 50$ Mpc/h. To measure
correlations of smaller amplitude, and thus on larger scales, we need
to have samples in which the typical fluctuation of the average
density is, at least, a factor ten smaller than the present one.

Note that for the case of MG samples, and specifically for VL2 and
VL3, the amplitude of the correlation functions is of the order of
$\overline{\sigma}$ up to $250$ Mpc/h.  We stress however that one
should also care about whether the property of self-averaging is
satisfied in these samples, and thus whether the determination of
average quantities gives a meaningful estimation of intrinsic
properties \citep{sdss_pnas,sdss_aea}.

\begin{figure}
\begin{center}
\includegraphics*[angle=0, width=0.5\textwidth]{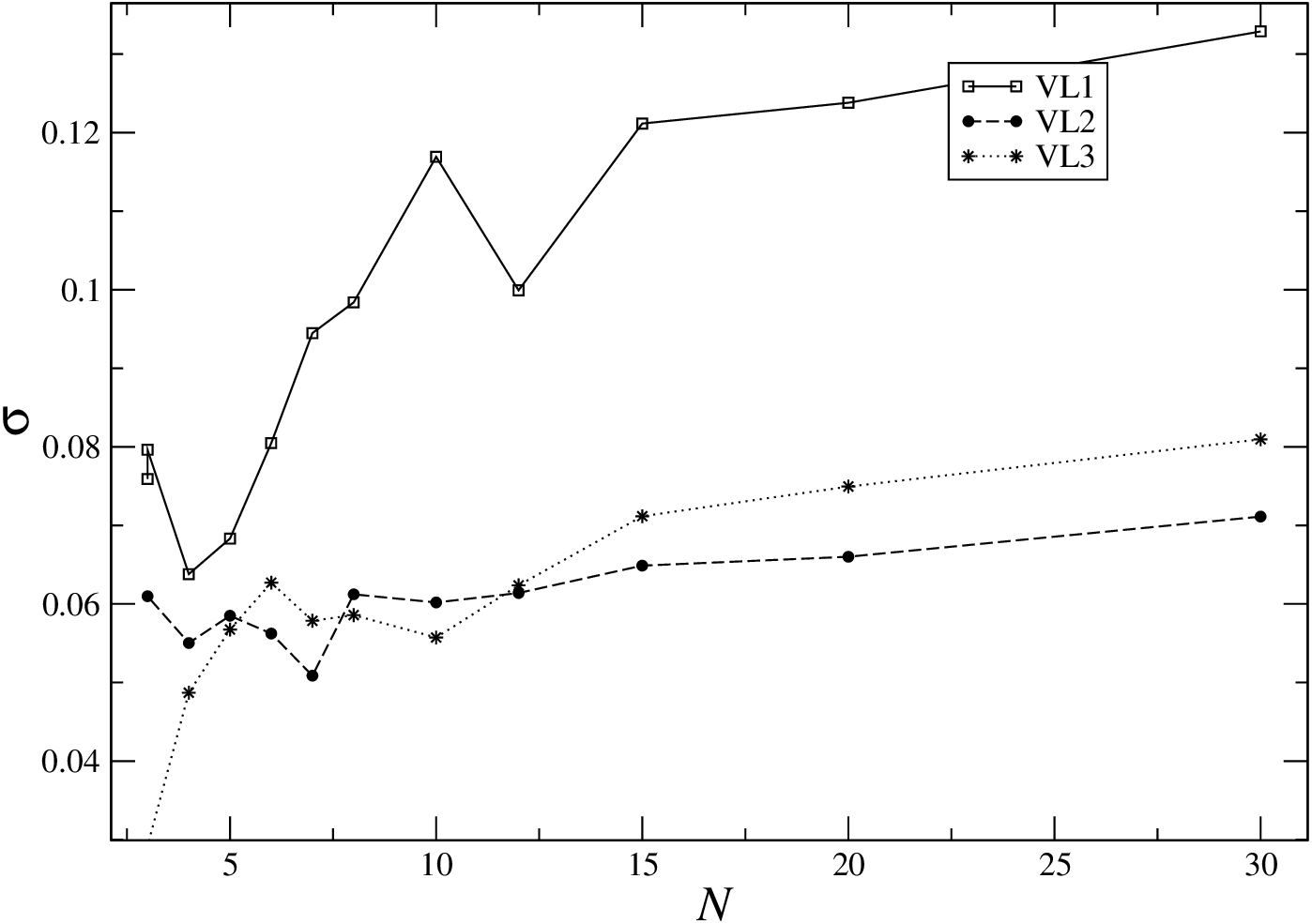} 
\end{center}
\caption{The same of Fig.\ref{flucts} but for the VL1, VL2 and VL3 case.} 
\label{flucts2}
\end{figure}

While the above argument about error propagation strictly applies when
we determine the correlation function by considering Eq.\ref{xi2}, we
show in what follows that the above estimation holds also in the case
of the DP and LS estimators.  To show this, let us now compute
statistical error bars in different way than by the JK method.

\subsection{Statistical errors} 

 The errors on the correlation function can be determined in various
 manners and the problem is to understand, in the case of the actual
 distribution, which methods gives the most reliable error estimation.
 To this aim, let us consider in more detail the computation of
 JK errors: in practice one takes almost fixed the sample
 density and computes the typical variation with respect to it.
 Indeed, we remind that each of the $N$ sub-fields used to in the
 JK estimation is equal the full sample without a small
 sub-field of angular area equal to $1/N$ of the full sample
 area. Therefore the different sub-fields are strongly overlapping: in
 the case in which large scale correlations are not negligible this
 method underestimates the errors in the correlation function
 estimation.

We find that in the LRG sample the variation of the sample density in
the $N=30$ sub-fields used to compute the JK errors is smaller,
i.e. $\overline{\sigma} \approx 5\cdot 10^{-3}$ than what is estimated
by computing the variance in {\it non-overlapping} sub-fields.  This
result does not show a particular dependence on the number of
sub-fields used as long as $N>10$.

Field-to-field errors can quantify,  volume-dependent systematic
effects due to large-scale variation of the sample density.  They can
be computed by dividing the sample into $N$ non-overlapping
sub-fields.  The correlation function can be estimated by
\be
\label{xiave}
\overline{ \xi(r)}  = \frac{1}{N} \sum_{i=1}^{N} 
\overline{\xi_i(r)}
\ee
and then the variance  is 
\be
\label{xisigma} 
\overline{\sigma_{FtF}^2 (r)} =  \sum_{i=1}^{N} \frac{(
\overline{\xi_i(r)} - \overline{ \xi(r)} )^2}{N-1} \;.
\ee

In Fig.\ref{errors} we show the behavior of the errors, in the LRG
sample, computed by Eq.\ref{jackerrors} and Eq.\ref{xisigma} and
considering  10, 20 and 30 fields.  One may note that (i) the
field-to-field error in larger than the signal for $r \gtapprox 50$
Mpc/h, i.e. for scales larger the amplitude of the estimated
correlation function is $\xi(r) \sim \overline{\sigma}$. (ii) The
field-to-field error is larger than the JK error on all scales
by about five times. 
 Note that the JK errors are similar to those derived by of
   \citet{cabre}. The field-to-field errors are much larger, and they
   could be over-estimates because the fields used are smaller than
   the full sample. To check whether this is the case we can vary the
   number of sub-fields used to estimate the field-to-field
   fluctuations as we did, for instance, to compute the typical rms
   fluctuation on the average density (see
   Figs.\ref{flucts}-\ref{flucts2}). Clearly by reducing the number of
   fields $N$ one has less determinations, while increasing $N$ one is
   finally dominated by shot noise. For $N$ in the range [10,30] we do
   not notice any clear decrease in the field-to-field errors. Our
   conclusion is therefore that the JK error is not the complete error
   but only the sampling error while the field-to-field fluctuations
   include the possible fluctuations due to the uncertainty on the
   sample density estimation and it and should be larger or equal than
   JK errors. 
An additional problem we consider in the next section, is whether the
statistical errors measured by considering non-overlapping fields are
able to take into account the whole uncertainty on the sample
  average, i.e. they can take into account the bias of the
  estimators.

 Note that the behavior of the correlation function in the MG VL2
  and VL3 samples on large enough scales, i.e. $r \approx 200$ Mpc/h,
  is the same when considering both JK and field-to-field errors,
  showing thus that there are positive correlations on scales larger
  than the cut-off on $r_c\approx 120$ Mpc/h predicted by the LCDM
  model without a statistical robust evidence of the $r_{bao}$ scale
  on $\approx 110$ Mpc/h.

\begin{figure}
\begin{center}
\includegraphics*[angle=0, width=0.5\textwidth]{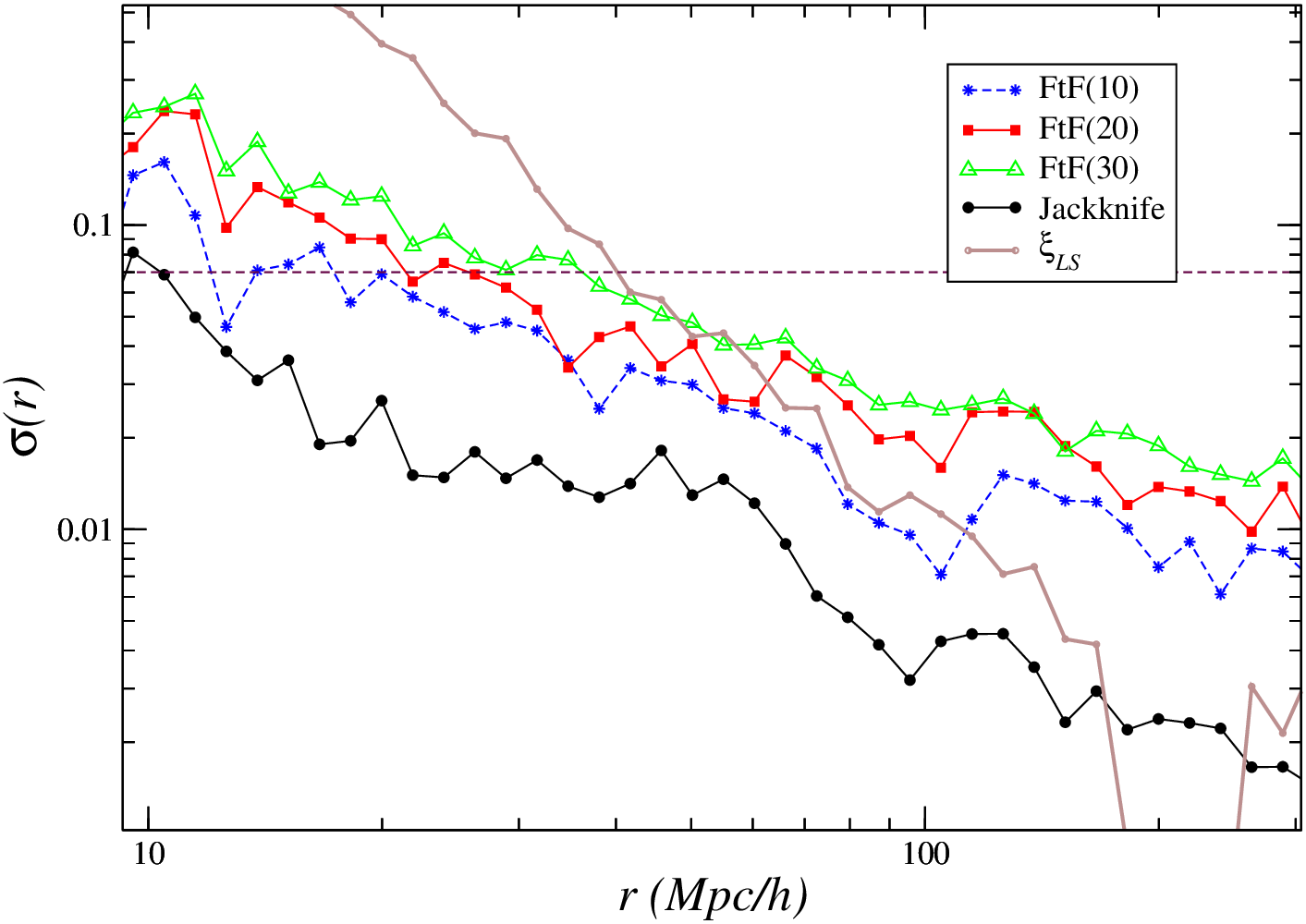}
\end{center}
\caption{Jack-knife errors, and field-to-field errors computed with
  different number of fields N=10,20,30 in the LRG sample. The
  solid line corresponds to the full-sample determination of the LS
  estimator.}
\label{errors}
\end{figure}

\subsection{Large scale  volume-dependent systematic effects}

The simple estimation of field-to-field errors allows one to overcome
the problem related to the JK method, in which the implicit assumption
is that correlations on the scale of the sample are
negligible. However the field-to-field method is not able to take into
account the full errors on the correlation function estimation.  This
is because the sample density is systematically different from the
ensemble average density when the correlation function is non zero at
large scales.  This introduces a well-known bias, i.e. a
  volume-dependent systematic effects. Let us discuss this further
effect.

Most of the literature on the correlation function measurements has
focused on the determination of the statistical errors
\citep{zehavi1,zehavi2,norberg2,lrg,norberg09} while little attention
has been devoted to the understanding of the distortions introduced by
volume-dependent systematic  effects.  These depend on the precise
type of estimator used, but they affect any estimator
$\overline{\xi(r;V)}$, in a finite sample of volume $V$, in some ways
at large enough scales \citep{cdm_theo}.

For instance, an important volume-dependent systematic effect is
related to the so-called integral constraint \citep{pee80} and can be
understood as follows.  The estimator $\overline{\xi(r;V)}$ measures
amplitude and shape of conditional correlations normalized to the
estimation of the sample mean instead to the ``true'' (ensemble or
infinite volume limit average) average density \citep{cdm_theo}. As
long as the ``true'' correlation function is different from zero
(e.g. in case of LCDM on all scales) any estimation of the average
density in a finite sample differs from the ``true'' value.  This
situation introduces a systematic distortion of $\overline{\xi(r;V)}$
with respect to $\xi(r)$ which, depending on the correlation
properties of the underlying distribution, is manifested in (i) an
overall difference in amplitude and (ii) a distortion of the shape for
$r \ltapprox V^{1/3}$ \citep{cdm_theo}.

In order words, only if the zero point of the correlation
  function is due to the boundary condition corresponding to the
  integral constraint, then this will be different for different
  sample sizes. If the zero-point is real, as it should be in a LCDM
  model, then it should not change from the sample to sample.

The definition of the range of scale in which this former effect
occurs, depends on the precise estimator used. For instance, in the
case of the full-shell (FS) estimator \citep{book,cdm_theo} and for a
spherical sample volume, our ignorance of the ``true'' average density
value is explicitly present in the condition that $\int_V
\overline{\xi(r,V)} (r) d^3r =0$, where the integral is performed over
the whole sample volume $V$. Note that this condition holds for any
$V$ and it forces the estimator to become negative even if the
``true'' $\xi(r)$ is always positive inside the given sample.  The
effect of this boundary condition is the following: as long as the
``true'' correlation function is positive, by enlarging the volume
size the change of sign occurs at larger and larger scales
\citep{cdm_theo}. This effect may very well explain the behavior found
in the MG VL samples discussed above, in which we noticed that the
transition scale changes from $r\approx 50$ Mpc/h for the smallest
sample to more than $250$ Mpc/h for the deepest sample we considered.
Note that if the ``true'' correlation function is negative, then the
distortion on large scales can be rather important \citep{cdm_theo}.

While for the FS estimator one can analytically calculate the scale at
which the systematic departure from the ``true'' shape occurs, for
more complex estimators based on pair-counting, like the LS one, it is
possible to understand only through numerical simulations the ways in
which this boundary condition affects the measured correlations. This
is the complication to be considered having the advantage that these
estimators can measure correlations on scales larger than those
sampled by the FS estimator \citep{cdm_theo}.  For pair-counting
estimators it has been numerically shown \citep{cdm_theo} that, when
fluctuations in the sample density are small enough, $r \propto
V^{1/3}$; the pre-factor of this proportionality depends on the type
of estimator and on the sample geometry.
However, we note that
 large scale fluctuations may alter this systematic behavior as
a function of the sample volume in a non trivial way (see e.g.,
\citet{2df_epl,2df_aea,sdss_pnas,sdss_aea}).

Note that the simple computation of how
the error in the average density propagates into the error on the
correlation function does not take explicitly 
into account of the situation in which the sample density 
itself can be a varying function of the sample size (the
interested reader to \citep{sdss_pnas,sdss_aea,2df_aea,2df_epl} for a
more complete discussion of this important point). 
Indeed, as mentioned above, the estimated sample average converges
to the asymptotic average density with a rate determined 
from the decaying of the two-point correlation function. When 
correlations are strong, there can be an important 
finite-volume dependence of the sample density,
resulting in a similar finite-size effects of
the two-point correlation function \citep{cdm_theo}.

\section{Theoretical implications} 
\label{theo}

To theoretically interpret these results it is necessary to take into
account an important complication which changes the predictions of
standard models described in the introduction. Indeed, these refer to
the whole matter density field (dark and luminous) while we observe
only a part of it in the form of luminous matter (i.e.  galaxies). The
relation between galaxy and dark matter distributions is usually
formulated in terms of bias: the latter represent a certain (physical)
sampling of the former.  There are two different relevant regimes. At
non-linear scales, where the distribution has strong clustering
characterized by non Gaussian fluctuations, this relation can be
studied only through numerical models \citep{springel,cronton}.
Instead, on scales where perturbations are small and clustering is in
the linear regime, there is a simple picture based on the threshold
sampling of a Gaussian random field \citep{kaiser}. In the former
  case one may derive analytically that the ``biased'' two-point
  correlation function is linearly amplified by threshold sampling
  \citep{kaiser}. This is found to occur also in the non-linear
regime but under different conditions, as shown by numerical N-body
simulations \citep{springel,cronton}: the effect of biasing is to
linearly amplify the correlation function, while the simple threshold
sampling of a Gaussian random field predicts a strongly scale
dependent amplification of the correlation function in the non-linear
regime \citep{bias1}.

Therefore the prediction of the non-linearity scale $r_0$ for the full
matter distribution (which, in current models, is $r_0 \approx 8$
Mpc/h) gives only an approximate estimate for that of galaxies of
different luminosity. Indeed this scale has been found to slightly
vary in N-body simulations \citep{springel}. On the other hand
  the scale $r_c$ is not affected by biasing for the simple reason
  that it is located, in current models, on about $r_c \approx 120$
  Mpc/h where fluctuations have low amplitude and thus where both
  biasing and gravitational clustering give rise to a linear
  amplification of the correlation function.  Hence, given that for
$r > r_c$ there are not positive correlations in the whole matter
density field, these will not be present in the galaxy distribution as
they cannot be generated by a biasing mechanism. Thus the length scale
$r_c$ is invariant with respect to biasing, i.e. it must be the same
for any class of objects as for the whole matter density field. It is
then a fundamental scale to be measured in the observed galaxy
distribution to verify the class of models characterized by the
Harrison-Zeldovich tail of the matter power spectrum. Finally the
third length scale in current models is the BAO scale, located at
$r_{bao} \ltapprox r_c$, and it is weakly affected by gravitational
evolution and biasing \citep{eh}.

That the scales $_{bao}$ and $r_c$ are invariant under biasing is
shown by the analysis of the N-body simulations provided by the Horizon project
\citep{horizon} where it is found that these are
the same for the whole matter distribution and for the sub-sample of
particles corresponding to the LRG  (see their Fig.5).


\section{Conclusions} 
\label{conclusions} 

In the newest SDSS samples it is possible to measure the correlation
function on $\sim 100$ Mpc/h scales both in the Main Galaxy (MG) and
in the Luminous Red Galaxy (LRG) samples.  We measured, in the former
case, positive correlations extending up to a factor two beyond the
scale $r_c\approx 120$ Mpc/h, at which in the LCDM model $\xi(r)$
should cross zero being negative on larger scales.  However in nearby
samples we measured that positive correlations are detectable only up
to $\sim 50$ Mpc/h.  Therefore we concluded that in these samples The
correlation function shows a rather different behavior from the LCDM
model prediction and that there is no statistical significant evidence
for the scale corresponding to the baryonic acoustic oscillations
(BAO).
Moreover we found that the estimated two-point correlation function in
different MG VL samples shows a clear dependence on the sample
volume. We concluded that the overall errors in the estimation of the
correlation function cannot be simply evaluated by the computation of
statistical error bars (e.g. JK) but they can only be studied
by making systematic tests in samples with different volumes.

In addition, we have shown that, in the LRG sample, the
uncertainty on the sample density estimation does not allow to
measure the correlation function on scales of the order of $\sim 100$
Mpc/h.  Rather it puts a upper limit to the estimation of correlations
at about $\sim 50$ Mpc/h.
More specifically the fluctuation on the estimation of the sample
density for the LRG sample is of the order of $8 \%$. This is, as we
have discussed, of the same order of the errors in the correlation
function. We have pointed out that in order to measure the small bump
in the correlation function associated with the BAO scale, one would
need samples in which the fluctuation on the estimated density is more
ten times lower than the value found in the LRG-DR7 sample.

For this reason we concluded that in the LRG sample there is no
statistical evidence for the BAO and that previous measurements
\citep{lrg,cabre,martinez} have underestimated the error bars in the
estimation of the correlation function and neglected the possible
  effect of the bias in the estimator. This is due to the fact that
they have measured statistical errors by means of the JK method. This
computes the sample variance by considering different samples which
are strongly overlapping. If large scale correlations are not
negligible, this method underestimates the errors in the correlation
function.  We have shown that a more reliable way to compute
statistical error bars is given by the simple estimation of
field-to-field fluctuations.  However, we have pointed that even this
method is not able to properly take into account overall
volume-dependent effects,  i.e. the estimator's bias, related to
our ignorance of the ensemble average density.

Determinations of correlations through the measurements of the galaxy
power spectrum \citep{cole} are affected by similar volume-dependent
systematic effects \citep{luca}. In addition one must take into
account that threshold sampling of a Gaussian field {\it does} change
the shape of power spectrum on large enough scales, i.e. on small
enough wave-numbers \citep{bias}.  A similar situation should occur in
the case of the halo models \citep{book}.

This situation represents an important challenge for models,
especially in view of the fact that galaxy distribution does not
present the negative correlations predicted by models up to scales
larger than $\sim 250$ Mpc/h. Our conclusion is that, in view of the
finite-volume effects, the estimation of correlations presented here
must be intended as a lower limit to the real correlations
characterizing the large scale distribution of galaxies.
Future surveys, like the extended SDSS III project \citep{Schlegel},
may allow us to study the behavior of the galaxy correlation function
on scales larger than those considered here.
To understand how volume-dependent systematic effects perturb
correlation measurements and to make tests on the volume stability of
statistical quantities it is necessary to consider a more complete
statistical analysis that focuses on conditional fluctuations
\citep{2df_epl,2df_aea,sdss_pnas,sdss_aea}.

\acknowledgements FSL is grateful to Andrea Gabrielli and Michael
Joyce for interesting discussions.  YVB thanks for partial support
from Russian Federation grants: Leading Scientific School 1318.2008.2
and RFBR 09-02-00143.  MLC was supported by the {\it Ram\'on y Cajal}
Program of the Spanish Science Ministry.  We thank an anonymous
  referee for a list of suggestions and criticisms that has allowed us
  to improve the presentation.  We acknowledge the use of the Sloan
  Digital Sky Survey data ({\tt http://www.sdss.org}) and of the NYU
  Value-Added Galaxy Catalog ({\tt http://ssds.physics.nyu.edu/}).


{}

\end{document}